\documentclass[aps,twocolumn,prl,showpacs,color,psfig,epsf]{revtex4}
\usepackage{amsmath}
\usepackage{color}
\usepackage{amsfonts}
\usepackage{epsf}
\usepackage{graphicx}
\begin{document}

\title{Colloids in active fluids: Anomalous micro-rheology and negative drag}

\author{G. Foffano$^1$, J. S. Lintuvuori$^1$, K. Stratford$^2$, M. E. Cates$^1$, D. Marenduzzo$^1$}
\affiliation{$^1$SUPA, School of Physics and Astronomy, University of Edinburgh, Mayfield Road, Edinburgh, EH9 3JZ, UK;\\
$^2$EPCC, School of Physics and Astronomy, University of Edinburgh, Mayfield Road, Edinburgh, EH9 3JZ, UK.}

\begin{abstract}
We simulate an experiment in which a colloidal probe is pulled through an active nematic fluid. We find that the drag on the particle is non-Stokesian (not proportional to its radius). Strikingly, a large enough particle in contractile fluid (such as an actomyosin gel) can show negative viscous drag in steady state: the particle moves in the opposite direction to the externally applied force. We explain this, and the qualitative trends seen in our simulations, in terms of the disruption of orientational order around the probe particle and the resulting modifications to the active stress.
\end{abstract}

\maketitle

Active particles take energy from their surroundings and convert this into mechanical work. Active fluids are suspensions of such particles in a Newtonian solvent, and they represent an interesting class of nonequilibrium soft matter~\cite{sriram_review,Ramaswamy_PRL_2004}. To lowest order, an active fluid may be modelled as a collection of force dipoles, exerted by the active particles, which quite often have orientational order (e.g., as a nematic phase).  These force dipoles are either contractile, when the forces are exerted ``inwards'' towards the centre of mass of each particle, or extensile, in the opposite case. Suspensions of bacteria such as {\it E.coli} are extensile fluids, while a dispersion of {\it Chlamydomonas} (algae) is contractile, as is the actomyosin gel which constitutes the cytoskeleton of eukaryotic cells~\cite{sriram_review}.

The continuous dipolar forcing present in active fluids profoundly affects their macroscopic properties. For instance, active nematics flow spontaneously in steady state in the absence of any external force, provided the activity level (dipole density) is high enough~\cite{spontaneousflow,spontaneousflowvoituriez}. Simulations~\cite{spontaneousflow,smfpre} show this flow to be chaotic, resembling the ``bacterial turbulence'' observed in concentrated films of {\it B. subtilis}~\cite{goldstein}. Furthermore, the bulk rheology of active fluids is strongly non-Newtonian~\cite{Ramaswamy_PRL_2004,kikuchi}. Theory predicts an increase (reduction) in the effective viscosity for contractile (extensile) fluids~\cite{Ramaswamy_PRL_2004}. These predictions were confirmed by simulations~\cite{fieldingPRL,Haines2011}, and also by experiments on {\it Chlamydomonas}~\cite{Rafai_PRL} and {\it E. coli}~\cite{Aranson}.
  
The {\it local} flow properties of active fluids also differ from their passive counterparts. These can be addressed by monitoring the dynamics of a probe particle in a ``micro-rheology" experiment. Such studies~\cite{Waigh} find marked violations of the fluctuation-dissipation theorem~\cite{lubensky,mizuno,fdt}, which in near-equilibrium systems links the decay of random fluctuations to the linear force response. Local flow of active fluids is of strong biophysical relevance: for instance, the cytoskeleton in moving cells is mainly subjected to localized cues rather than global forces, and its response to these may be crucial to cell motility~\cite{Bray}.

Here we address a very basic issue of the local flow response in active fluids. We ask: what happens when a passive particle of radius $R$ is dragged through an active nematic with a force $F$? This represents perhaps the simplest micro-rheological experiment possible on an active fluid.
We find by simulations that this simple experiment should lead to some very interesting results. First, the drag coefficient
 $\xi=F/v$ (with $v$ a steady particle speed) exhibits a strongly nonlinear dependence on radius, in violation of Stokes' law. This is especially noticeable in the contractile case. We explain these violations in terms of the deformation of the orientational order in the active fluid around the probe particle, and present a simple scaling theory for the balance between active and viscous forces. Second, and strikingly, our theory and simulations show that in a contractile fluid the colloidal probe, if large enough, should move steadily in the direction {\em opposite} to the applied force. That is, we find a stable steady state with a negative drag coefficient. We also simulate a transient (force reversal) experiment that probes further the remarkable physics in this regime. 

{\em Numerical model:}
In the continuum limit, the hydrodynamics of an active nematic fluid can be described by a set of continuum equations~\cite{sriram_review,Ramaswamy_PRL_2004} that govern the time evolution of the velocity field $u_\alpha$ and of a (traceless, symmetric) tensor order parameter $Q_{\alpha\beta}$. The latter describes the orientational order of the active particles (whether bacteria, algae, or cytoskeletal filaments) which usually have a rod-like shape and are thus capable of nematic alignment~\cite{sriram_review}. Without activity, nematics are described by a Landau -- de Gennes free-energy density ${\cal F} = F(Q_{\alpha\beta}) + K(\partial_{\beta}Q_{\alpha \beta})^2/2$, with \begin{equation}\label{eq:fed_bulk}
F(Q_{\alpha\beta}) = \left(1-\frac{\gamma}{3}\right)\frac{Q_{\alpha \beta}^2}{2}-\frac{\gamma}{3}Q_{\alpha \beta}Q_{\beta \gamma}Q_{\gamma \alpha} + \frac{\gamma}{4}(Q_{\alpha \beta}^2)^2
\end{equation}
where indices denote Cartesian coordinates, summation over repeated indices is implied, $\gamma$ controls the magnitude of nematic order, and $K$ is an elastic constant.  

The hydrodynamic equation for the evolution of the order parameter is:
$D_t Q_{\alpha \beta} = \Gamma H_{\alpha \beta}$,
with $D_t$ a material derivative describing advection by the fluid  velocity $u_\alpha$, and rotation/stretch by flow gradients (see \cite{spontaneousflow}). The molecular field is $H_{\alpha \beta}= -{\delta 
{\cal F} / \delta Q_{\alpha \beta}} + (\delta_{\alpha \beta}/3) {\mbox {\rm Tr}}({\delta {\cal F} / \delta Q_{\alpha \beta}})$ 
and $\Gamma$ is an inverse rotational friction. 
The fluid velocity obeys $\partial_\alpha u_\alpha = 0$, and also the Navier-Stokes equation, in which a passive $Q_{\alpha\beta}$-dependent thermodynamic stress enters~\cite{spontaneousflow}. The active force dipoles then create a further stress,
\begin{equation}\label{activestress}
\Pi_{\alpha\beta}= 
-\zeta \left(Q_{\alpha \beta}+\frac{1}{3}\delta_{\alpha \beta}\right),
\end{equation}
where $\zeta$ is the activity parameter that sets the dipolar force density~\cite{Ramaswamy_PRL_2004}. Note that $\zeta<0$ for contractile fluids and $\zeta>0$ for extensile ones.
Within our hybrid numerical scheme, we solve the Navier-Stokes equation via lattice Boltzmann, and the equation for the order parameter via finite difference~\cite{Juho}. Periodic boundary conditions are deployed throughout.

We introduce a spherical colloidal probe by a standard method of bounce-back on links~\cite{Nguyen} which provides a no-slip boundary condition for the velocity field at the solid surface. Order parameter variations create additional elastic and active forces on the particle which are computed by integrating the total stress tensor over its surface. There we impose planar anchoring of the nematic director, which as usual is a headless unit vector $n_\alpha$ oriented along the major principal axis of $Q_{\alpha\beta}$. Planar anchoring is imposed via an additional quadratic term in the free energy (see~\cite{galatola,Juho2} for details).

Below we give our results in simulation units~\cite{simulation-units}. To convert them into physical ones, relevant for instance to a contractile actomyosin solution, we can assume a value of $K$ of 1.25 nN, and a rotational viscosity of 1.0 kPa/s. (These values hold for typical cytoskeletal gels~\cite{mogilner}.) In this way, the simulation units for force, length and time can be mapped onto 25 nN, 0.5 $\mu$m and 10 ms respectively. Note that the same equations also apply to an extensile bacterial fluid, but the mapping to physical units in that case leads to very different values~\cite{spontaneousflow}.  Our model also neglects motility which is important in bacterial fluids, where it naturally leads to density inhomogeneities~\cite{spontaneousflow,giomimarchetti}.

\begin{figure}\begin{center}
\includegraphics[width=0.5\textwidth]{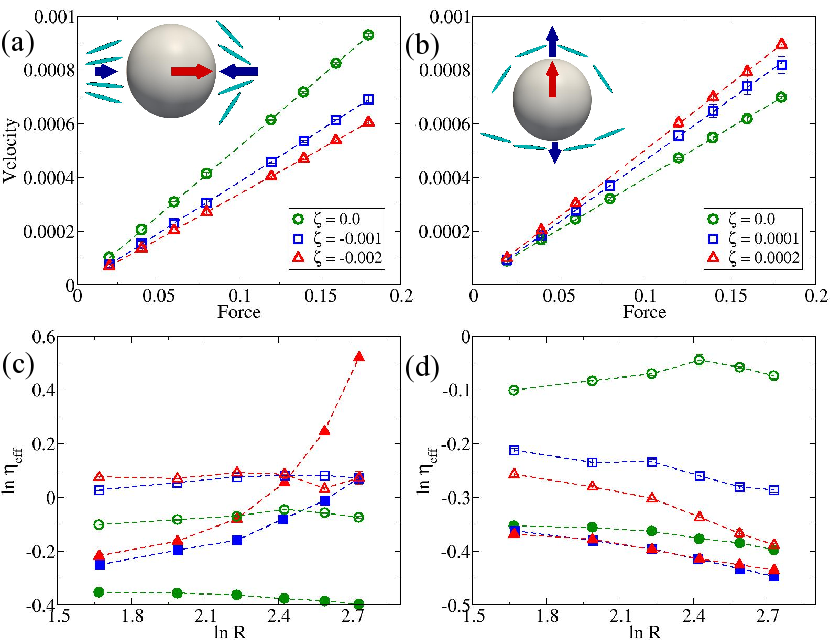}
\label{fig:no_walls}
\caption{(Color online) (a,b) Plots of $v(F)$ curves for colloids of fixed radius $R=11.3$, at various $\zeta$ (see legend) in a contractile (a) and an extensile (b) active nematic, dragged respectively parallel and perpendicular to the far-field director. (c,d) Dependence of $\eta_{\rm eff}$, measured according to Eq.~\ref{effective_viscosity}, for both (c) contractile and (d) extensile cases. Filled symbols refer to an external force parallel to the far-field director, while empty symbols are used for orthogonal drag. Different symbols (colours) refer to different activities (see legend in (a) and (b)). In (a) and (b), insets, we sketch the director field and active force directions for a colloid pulled (a) in a contractile active fluid along the far-field director, and (b) in an extensile fluid perpendicular to it.}
\end{center}
\end{figure}

{\em Results:} We now discuss the result of our micro-rheological simulation in which a colloid of radius $R$ is pulled through an active fluid, either contractile or extensile. 
The external force $F$ was directed either along or perpendicular to the far-field nematic director. For the values of $R$ chosen, plots of the steady state velocity $v$ (measured along $F$) versus $F$ (see Fig.~1a,b) show a well-defined linear regime at small external force. In all our simulations $\zeta$ is kept small enough such that no spontaneous flow arises in the absence of the probe particle~\cite{notespontaneousflow}.
Figs.~1a,b show that the linear drag coefficient $\xi_0 = F/v|_{F\to 0}$ increases with activity for a contractile fluid, and decreases for an extensile one (within the parameter range that we explored). Thus a contractile fluid opposes motion more strongly than its passive counterpart; an extensile fluid less. So far, this is in line with the respective increase (contractile) and decrease (extensile) of bulk fluid viscosities mentioned above~\cite{yield}. 
 
By analogy with the formula for Stokes drag on a probe of radius $R$ in a passive fluid of viscosity $\eta$, we can define an effective viscosity for the probe motion as 
\begin{equation}\label{effective_viscosity}
\eta_{\rm eff}(R)=\xi_0/(6\pi R).
\end{equation}
This is plotted as a function of $R$ in Figs.~1c,d after correction for periodic boundary effects~\cite{note_hasimoto}.
If Stokes' law {\it did} hold, the effective viscosity should be independent of $R$, and we see that this is indeed the case for a passive nematic ($\zeta=0$, open and filled circles in Fig. 1c,d). When activity is switched on, this picture changes dramatically and the drag coefficient becomes strongly non-Stokesian. For a contractile fluid, this anomaly is most apparent when pulling along the director field, where the effective viscosity increases sharply with the radius. In the extensile case the most markedly anomalous response is instead obtained when the particle is pulled {\it perpendicularly} to the far-field director, and the effective viscosity this time drops with size. 

These results show that for active fluids
microrheology experiments do not simply probe the fluid, but measure a scale-dependent property of the probe and surrounding fluid in combination. Although microrheology is thereby disconnected from bulk rheology, such experiments may offer essential insights into the physical response of (for instance) an actomyosin gel on the length scale relevant to transport of organelles or other subcellular objects~\cite{Bray}.

We can qualitatively explain the anomalous drag $\xi_0(R,\zeta)$
in Fig.~1c via the following argument (sketched in Fig.~1a, inset). As the colloidal probe is pulled through the contractile fluid, it deforms the director field. With the pulling force oriented parallel to the far-field director, planar anchoring requires an elastic splay of the director locally. Because of the fluid flow created by particle displacement, this splay will be stronger in front of the moving particle than behind it. Through its effect on the active stress, the splay will result in a large force opposing the motion at the front, and in a smaller force favoring it at the rear. The net effect is to slow the particle down: $\xi_0$ is increased. The same argument holds for pulling perpendicular to the director, but here is much smaller.
We attribute the different magnitude of the effect in the two directions to the fact that contractile activity enhances the splay response (leading at large wavelengths to an instability involving splay but not bend~\cite{sriram_review}). 
One can likewise understand the results in Fig.~1d for an extensile fluid, by sketching the expected nematic deformation in front and at the rear of the particle (see inset in Fig.~1b). Now the incipient instability is towards bending, so pulling perpendicularly to the director gives the larger effect. Moreover, the active stress on the front of the particle now favors motion and the smaller one at the back opposes it: so drag is reduced. These arguments suggest that the $v(F)$ curves should strongly depend on the anchoring conditions at the particle surface, and we have confirmed this numerically (data not shown).

Building more quantitatively on these arguments, we now estimate the extra force acting on a moving particle arising from the active contribution to the stress (integrated over the colloidal surface). This depends on the surrounding flow field and nematic deformation, details of which can only be computed numerically. Nonetheless, we can argue that activity leads to an additional force on the particle which, on dimensional grounds, equals $A(F) \zeta R^2$. To estimate $A$, we note that for $F=0$ the particle is immobile, so that the extra force should also vanish; $A(0)=0$. To first order, dimensional analysis now suggests $A\sim c F/K$ with $c$ a positive dimensionless number of order unity~\cite{nonlin}. (Our previous arguments show that $A/F$ is positive for both signs of $\zeta$.) 
The force balance for the moving colloid in steady state is then given by $F+c \zeta FR^2/K=6\pi \tilde{\eta} R v$, where $\tilde{\eta}$ is the passive nematic viscosity, which may depend on pulling direction. This equation together with Eq.~\ref{effective_viscosity} leads to the following prediction for the effective viscosity
\begin{equation}\label{theory}
\frac{1}{\eta_{\rm eff}} = \frac{1}{\tilde{\eta}}\left(1+\frac{c \zeta R^2}{K} \right)
\end{equation}

This provides a surprisingly robust explanation for the size dependence of the drag coefficient: Fig.~2a shows that for a colloid in a contractile fluid, plotting $1/{\eta_{\rm eff}}$ versus $R^2$ yields a straight line as predicted by Eq.~\ref{theory}. By fitting our numerical data, we also obtain $c\sim 0.06$ for parallel pulling independent of $\zeta$, which further validates our approximate theory. However Eq.~\ref{theory} works less well in the extensile case. There the effect is smaller, and passive contributions to the stress tensor possibly play a more significant role. 

\begin{figure}
\includegraphics[width=0.5\textwidth]{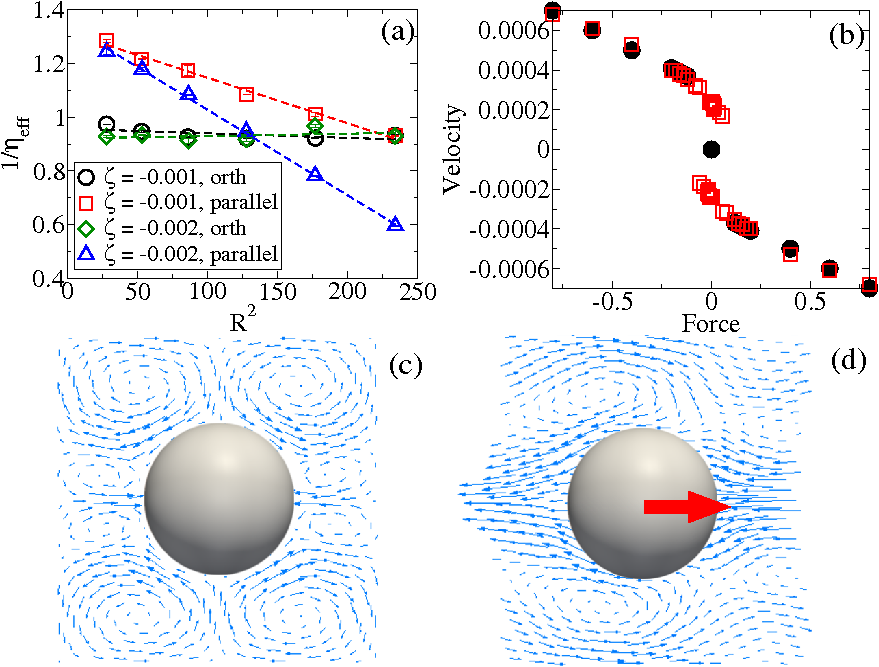}
\label{fig:fit}
\caption{(Color online) (a) Plot of $1/\eta_{\rm eff}$ on $R^2$, yielding an approximately linear dependence in agreement with Eq.~\ref{theory}. (b) $v(F)$ curve for a case with negative drag ($R = 30,\zeta = -0.002$). Filled circles and open squares refer to simulations in which the force is applied to a quiescent and moving colloid respectively.
The velocity fields in (c) and (d) correspond to the case of a quiescent particle (c) and of a particle moving to the left, opposite to the applied force (large arrow) (d).}
\end{figure}

Our simplified theory in Eq.~\ref{theory} formally predicts that when the dimensionless quantity ${c\zeta R^2}/{K}<-1$, $\eta_{\rm eff}<0$, giving a negative drag coefficient ($\xi = F/v<0$). However, one might reasonably expect that before this regime is reached our leading-order expression $A\sim cF/K$ would break down and some different physics would come into play, so that $\xi$ remains positive. Remarkably though, this is not the case. Although positive $\xi(F)$ appears to be restored at extremely small forces, for modest but finite $F$, we robustly find that the particle moves in the opposite direction to the externally applied force. A steady-state $v(F)$ curve showing this effect for ${c\zeta R^2}/{K} \sim 2.2$ is in Fig.~2b.
This reverse-sigmoidal curve shows bistability at small forces, with two stable branches each of negative drag, $\xi = F/v<0$, and also negative slope, $dF/dV<0$. Interestingly, there seems to be no simple relation between negative local drag in contractile fluids and the homogeneous bulk rheology of active fluids. In the latter, a negative ratio of stress to strain rate can arise, but only in the {\em extensile} case~\cite{spontaneousflow}. Moreover, the resulting flows are generically unstable~\cite{smfpre}. 

To shed more light on the nature of the negative drag state in contractile fluids, we plot in Figs.~2c,d the active fluid flow around a stationary, unforced probe particle and that for one moving against the external force. For the static particle, a symmetric pattern of eddies is created by contractile forces arising from the deformation of the director (horizontal in far field), which splays around the sides of the probe and bends at its top and bottom. In the negative drag regime with the external force acting to the right, the eddies become asymmetric, creating a packet of left-moving fluid. The resulting leftward advective velocity of the particle exceeds its rightward speed relative to the local fluid packet, giving an overall leftward motion in the lab frame. 

We have examined how this remarkable steady-state flow pattern is reached dynamically. On introducing a rightward force, the particle first moves rightward (see Fig.~3) creating strong elastic distortions on its leading (right-hand) side. Such deformations then establish the leftward moving fluid packet (as in Fig.~2d), which sets up deformation in the surrounding director field to finally create the steady state travelling flow packet with the particle, as well as the fluid, moving leftward.  
We emphasize that this spontaneously moving state is quite distinct from the bulk spontaneous flows known to occur in active fluids when activity exceeds a system-size dependent threshold \cite{spontaneousflowvoituriez}. Because we are below that threshold, in our simulations the fluid velocity is essentially localized in the neighborhood of the probe particle. 

\begin{figure}
\includegraphics[width=0.45\textwidth]{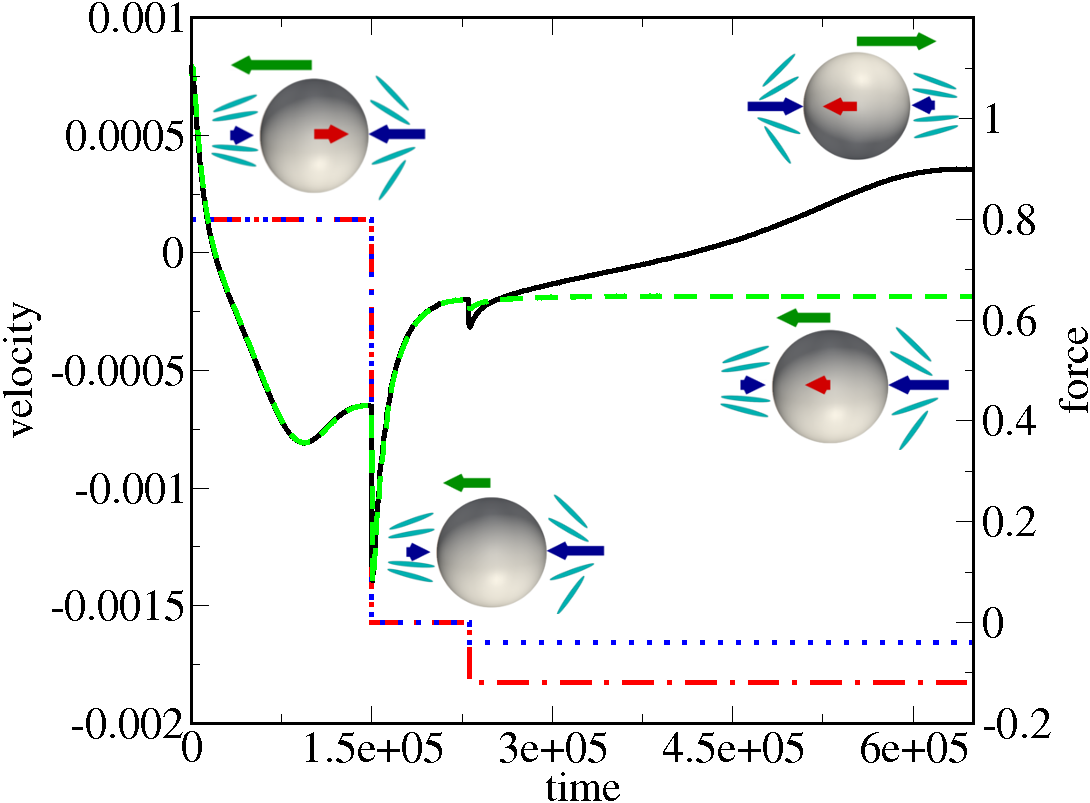}
\label{fig:dynamics}
\caption{(Color online.) Dependence of the external force (step functions, axis on the right) and of the particle velocity (axis on the left) on time for two different 'experiments'. In both cases a force $F=0.8$ is initially applied to the particle and turned off. Then a smaller force is applied along the direction of particle motion: the blue dotted line and the red dot-dashed one give the values of the external force in the two cases. The correspondent particle velocities are given by the black solid and green dashed lines respectively. The dynamics is discussed in the text. In the sketches arrows on the fluid refer to the active force direction, the arrow on the colloid represents the external force, while that above the particle shows the actual direction of motion.}
\end{figure}

A further exploration of the bistable negative-drag region at small $F$ is presented in Fig.~3. Here, we start with a static particle, of radius large enough to give negative drag ($c\zeta R^2/K \sim 2.2$) and first pull it to the right. The transient response was already described above, and leads finally to steady leftward
motion. We now reduce $F$ to zero: the particle keeps moving to the left, but slows down. We next start pulling the particle with a tiny leftward force: the probe continues to move leftward, albeit even more slowly than before. If this last stage is repeated with a much larger leftward force, the active flow and deformation pattern around the colloid dynamically reconstructs itself with the opposite sense, leading to a rightward probe velocity and once more to a negative drag. 
This  sequence of states can be viewed as a trajectory on the $v(F)$ curve of  Fig.~2b. The initial large rightward force puts the particle on the lower branch of the curve to the right of the vertical axis. As the force is removed and then applied to the left, the system ascends this lower branch. Where that branch ends, the system must jump to the upper branch.

{\em Conclusion:} we have simulated numerically a simple microrheological experiment, which should be realizable in the laboratory (with some caveats, noted below). In this experiment a colloidal probe is pulled through an active nematic fluid. We have shown that the activity leads to a non-Stokesian drag force that increases approximately quadratically with particle size. This behavior creates a new regime, arising in a contractile fluid such as an actomyosin gel, at large values of a dimensionless parameter ${c \zeta R^2}/{K}$. Strikingly, in this regime the colloidal probe is predicted to move against the driving force to create a stable steady state of negative drag. This contrasts with the bulk behavior of active fluids where negative ratios of stress to strain rate can arise in principle, but are unstable, and expected only for the opposite sign of activity (extensile rather than contractile)~\cite{spontaneousflow}.

Stable negative drag is counterintuitive but no physical law prevents it in active systems. Related phenomena have been reported for an ensemble of molecular motors with load-accelerated dissociation~\cite{negative_drag_howard}, in axon mechanics \cite{bernal}, for filament fluctuations in active media \cite{kikuchi}, and also in the upstream migration against a flow field of slime-mold cells~\cite{dicty} and bacteria \cite{nash}. In our context, the anomalous (ultimately negative) drag stems from the active stress, Eq.~\ref{activestress}, that emerges from a well-accepted coarse-grained description of activity in systems such as cytoskeletal gels~\cite{sriram_review}. 
Finally, an important requirement for
any experimental test of our predictions with actomyosin is that issues arising from macroscopic network clustering should be avoided~\cite{silva}. This might be feasible either through a suitable choice of parameters such as myosin concentration etc., or by working in metastable uniform networks. However, such experiments may well prove to be challenging.

We acknowledge EU network ITN-COMPLOIDS, FP7-234810, and EPSRC EP/J007404/1, for funding. MEC is funded by the Royal Society. 



\begin{thebibliography}{99}
\bibitem{simulation-units} Parameters (in simulation units) were: 
$K = 0.05,~\xi=0.7,~\gamma = 3.0$ and $\Gamma =0.5$. 
We used (periodic) cubic 
simulation box of volume $V=L^3=128^3$ and probe particles of various radii as indicated in the Figures. The strength of the planar anchoring was $W=0.05$.
\bibitem{yield} However, for contractile gels there is a bulk yield stress, whereas there is no sign of a corresponding ``yield force'' below which the probe cannot move in Fig.~1a. See \cite{fieldingPRL}.
\bibitem{nonlin} Note that any dependence of $A$ on $F$ other than the linear one we assume would lead either to the absence of any linear regime in the micro-rheology (if $A(F)$ is superlinear), or to an effective viscosity which is identical to the passive one (if sublinear). Both outcomes are ruled out by our data.
\end{thebibliography}
\end{document}